\documentclass[preprint,aps,draft,amsmath,amssymb,showpacs]{revtex4}
\usepackage{graphicx}
\usepackage{epsfig}
\usepackage{dcolumn}
\usepackage{bm}
\def\nA{nucleon-nucleus\ }
\def\pA{proton-nucleus\ }
\def\aA{$\alpha$-nucleus\ }
\def\AA{nucleus-nucleus\ }
\def\pn{$(p,n)$\ }
\begin{document}
\preprint{Draft for Phys. Rev. C}
\title{Neutron transition strengths of 2$^+_1$ states in the neutron rich Oxygen
isotopes determined from inelastic proton scattering}
\author{Nguyen Dang Chien$^{1,2}$}
\author{Dao T. Khoa$^{1}$}\email{khoa@vaec.gov.vn}
\affiliation{\mbox{$^1$Institute for Nuclear Science {\rm \&} Technique, VAEC}
\mbox{179 Hoang Quoc Viet Rd., Nghia Do, Hanoi, Vietnam.} \\
 $^{2}$ Department of Physics, Dalat University \\
 \mbox{1 Phu Dong Thien Vuong Str., Dalat, Vietnam.}}
\date{\today}
\begin{abstract}
A coupled-channel analysis of the $^{18,20,22}$O$(p,p')$ data has been performed
to determine the neutron transition strengths of 2$^+_1$ states in Oxygen
targets, using the microscopic optical potential and inelastic form factor
calculated in the folding model. A complex density- and \emph{isospin} dependent
version of the CDM3Y6 interaction was constructed, based on the
Brueckner-Hatree-Fock calculation of nuclear matter, for the folding model
input. Given an accurate isovector density dependence of the CDM3Y6 interaction,
the isoscalar ($\delta_0$) and isovector ($\delta_1$) deformation lengths of
2$^+_1$ states in $^{18,20,22}$O have been extracted from the folding model
analysis of the $(p,p')$ data. A specific $N$-dependence of $\delta_0$ and
$\delta_1$ has been established which can be linked to the neutron shell closure
occurring at $N$ approaching 16. The strongest isovector deformation was found
for 2$^+_1$ state in $^{20}$O, with $\delta_1$ about 2.5 times larger than
$\delta_0$, which indicates a strong core polarization by the valence neutrons
in $^{20}$O. The ratios of the neutron/proton transition matrix elements
($M_n/M_p$) determined for 2$^+_1$ states in $^{18,20}$O have been compared to
those deduced from the mirror symmetry, using the measured $B(E2)$ values of
2$^+_1$ states in the proton rich $^{18}$Ne and $^{20}$Mg nuclei, to discuss the
isospin impurity in the $2^+_1$ excitation of the $A=18,T=1$ and $A=20,T=2$
isobars.

\end{abstract}
 \maketitle

\section{Introduction}
Inelastic proton scattering has been used over decades as a very efficient tool
to yield the nuclear structure information. In difference from the
electromagnetic probes, protons interact strongly with both protons and neutrons
in the target nucleus, and the neutron and proton transition strengths of a
nuclear excitation could be reliably deduced from the $(p,p')$ measurement, in
terms of the neutron $M_n$ and $M_p$ matrix elements \cite{Be83}. The knowledge
of $M_n$ and $M_p$ can shed light into the relative contributions by the valence
nucleons and the core to the nuclear excitation, and hence, provides important
information on the core polarization by the valence nucleons which could
eventually lead to changes in the shell structure. This topic has recently
become of significant interest in the experimental studies with radioactive
beams where the inelastic proton scattering can be accurately measured, in the
inverse kinematics, for the short-lived unstable nuclei (see, e.g.,
Refs.~\cite{Je99,El00,Ct01,Be06,Iw08} for the $(p,p')$ measurements with the
unstable Oxygen isotopes). With large neutron (or proton) excess in the unstable
nuclei, such proton scattering data provide also a vital information for
studying the isospin effects in the \pA interaction. Although the isospin
dependence of the nucleon optical potential (OP), known by now as Lane potential
\cite{La62}, has been studied since a long time, few attempts were made to study
the isospin dependence of the transition potential or form factor (FF) for
\emph{inelastic} scattering. As neutron and proton contribute differently to the
nuclear excitation, the inelastic scattering FF contains also an isospin
dependence \cite{Gr80} which determines the degree of the \emph{ isovector}
mixing in the inelastic scattering channel that induces the excitation
\cite{Kh03}.

In general, the isospin-dependent part of the \nA OP is proportional to the
product of the projectile and target isospins, and the total OP can be written
in terms of the \emph{isoscalar} (IS) and \emph{isovector} (IV) components
\cite{La62} as
\begin{equation}
 U(R)=U_0(R)\pm\varepsilon U_1(R), \ \varepsilon=(N-Z)/A, \label{e1}
\end{equation}
where the + sign pertains to incident neutron and - sign to incident proton. The
strength of the Lane potential $U_1$ is known from $(p,p)$ and $(n,n)$ elastic
scattering and $(p,n)$ reactions studies, to be around 30-40\% of the $U_0$
strength. In the first order of the collective model, inelastic \nA scattering
cross section can be reasonably described, in the distorted-wave Born
approximation (DWBA) or coupled channel (CC) formalism, with the inelastic form
factor $F$ given by `deforming' the optical potential (\ref{e1}) as
\begin{equation}
 F(R)=\delta\frac{dU(R)}{dR}=\delta_0\frac{dU_0(R)}{dR}\pm\varepsilon\
 \delta_1\frac{dU_1(R)}{dR}. \label{e2}
\end{equation}
The explicit knowledge of the deformation lengths $\delta_0$ and $\delta_1$
would give us important structure information about the IS and IV transition
strengths of the nuclear excitation under study. There are only two types of
experiment that might allow one to determine $\delta_0$ and $\delta_1$ using
prescription (\ref{e2}):

i) Charge exchange \pn reaction leading to the \emph{excited} isobar analog
state. It was shown, however, that the calculated inelastic \pn cross sections
were insensitive to $\delta_1$ due to the dominance of two-step process
\cite{Fi79}.

ii) Another way is to extract $\delta_{0(1)}$ from the $(p,p')$ and $(n,n')$
data measured at about the same energy for the same excited state of the target
\cite{Fi79,Gr80}. Since $\varepsilon U_1/U_0$ is only about few percent, the
uncertainty of this method can be quite large. Moreover, it remains technically
not feasible to perform simultaneously $(p,p')$ and $(n,n')$ measurements in the
inverse kinematics for unstable nuclei.

From a theoretical point of view, the form factor (\ref{e2}) has been shown to
have inaccurate radial shape which tends to underestimate the transition
strength, especially, for high-multipole excitations induced by inelastic
heavy-ion scattering \cite{Be95,Kh00}. As an alternative, a compact approach
based on the folding model has been suggested in Ref.~\cite{Kh03} for the
determination of the IS and IV transition strengths of the ($\Delta S=\Delta
T=0$) nuclear excitations induced by inelastic proton scattering. This compact
folding approach was used with some success in the DWBA analyses of the
$^{30-40}$S$(p,p')$ and $^{18-22}$O$(p,p')$ data measured in the inverse
kinematics, to determine $\delta_0$ and $\delta_1$ for the 2$^+_1$ states in the
Sulfur and Oxygen isotopes under study \cite{Kh03,Kh07a}. We recall that the
basic inputs for such a folding + DWBA analysis are the effective $NN$
interaction between the incident proton and those bound in the target, and the
transition densities of the nuclear excitation. Consequently, for a carefully
chosen model of the nuclear transition densities, the more accurate the choice
of the effective $NN$ interaction the more reliable the deduced $\delta_0$ and
$\delta_1$ deformation lengths.

Our first folding model analysis of the $^{18,20}$O$(p,p')$ data \cite{Kh03} has
used a well-tested CDM3Y6 density dependent interaction \cite{Kh97} to construct
the \pA OP and inelastic FF. For simplicity, the density dependence of the
isovector part of the CDM3Y6 interaction has been assumed in Ref.~\cite{Kh03},
following a Hartree-Fock (HF) study of asymmetric nuclear matter \cite{Kh96}, to
be the same as that of the isoscalar part. As a result, quite a strong IV mixing
was found for 2$^+_1$ states in $^{18,20}$O, with the ratio of neutron/proton
transition matrix elements $M_n/M_p\simeq 4.2$ for $^{20}$O. Although that value
agrees fairly with previous estimates \cite{Je99,El00,Ct01} within the limits of
experimental errors, a recent measurement of the Coulomb excitation of $^{20}$Mg
\cite{Iw08} has revealed that the $M_n/M_p$ is only around 2.5 for 2$^+_1$ state
in $^{20}$O, if one assumes the $M_p$ moment of 2$^+_1$ state in $^{20}$Mg equal
the $M_n$ moment of 2$^+_1$ state in $^{20}$O based on the isospin symmetry. A
question was raised in Ref.~\cite{Iw08} whether such a discrepancy is due to the
inaccuracy of the $(p,p')$ analysis method of Ref.~\cite{Kh03} or the isospin
impurity in the 2$^+_1$ excitation of $^{20}$Mg and $^{20}$O.

In the mean time, the isovector density dependence of the CDM3Y6 interaction has
been carefully probed in the CC analysis of the \pn reactions exciting the 0$^+$
isobaric analog states of $^{6}$He \cite{Kh05} and other medium-mass nuclei
including $^{208}$Pb \cite{Kh07b}, where the isovector coupling was used to
explicitly link the isovector part of the nucleon OP to the cross section of \pn
reaction. In particular, a complex isovector density dependence of the CDM3Y6
interaction has been constructed based on the microscopic Brueckner-Hatree-Fock
calculation of nuclear matter \cite{Je77} by Jeukenne, Lejeune and Mahaux (JLM)
before being used as folding input. The main conclusion drawn from the results
of Refs.~\cite{Kh05,Kh07b} is that the strength the isovector density dependence
of the CDM3Y6 interaction, even after it was fine tuned against the JLM results,
is somewhat weak compared to the empirical isovector strength implied by the \pn
data. As a result, a renormalization of the (real) isovector density dependence
of the CDM3Y6 interaction by a factor of 1.2 - 1.3 was found \cite{Kh07b}
necessary to account for the measured \pn cross sections. Such an enhancement of
the isovector density dependence of the CDM3Y6 interaction was also shown
\cite{Kh05,Kh07b} necessary for a good HF description of the nuclear matter
symmetry energy compared to the empirical estimates.

Since a realistic isospin dependence of the effective $NN$ interaction is vital
for the determination of the IS and IV deformation lengths (or $M_n$ and $M_p$
moments), a \emph{revised} folding model analysis of the $^{18,20,22}$O$(p,p')$
data is necessary for a more definitive conclusion on the neutron transition
strength of 2$^+_1$ states in $^{18,20,22}$O isotopes. After a brief overview of
the theoretical formalism in Sec.~\ref{sec2}, the results of the folding + CC
analysis of the $^{18,20,22}$O$(p,p')$ data are presented in Sec.~\ref{sec3} and
the main conclusions are given in the Summary.

\section{General formalism}
\label{sec2}
\subsection{Nuclear densities, isoscalar and isovector deformations}
We describe here briefly the method suggested first in Ref.~\cite{Kh03} to link
the deformation of an excited nucleus and the corresponding transition density
based on a \emph{collective model} treatment. As the nuclear deformation is
associated with the `deformed' shape of excited nucleus, instead of `deforming'
the optical potential (\ref{e2}), one can build up the proton and neutron
transition densities of a $2^{\lambda}$-pole excitation ($\lambda\ge 2$) using
the so-called Bohr-Mottelson (BM) prescription \cite{Bo75} separately for
protons and neutrons
\begin{equation}
 \rho^\tau_\lambda(r)=-\delta_\tau\frac{d\rho^\tau_{\rm g.s.}(r)}{dr},\ {\rm with}
 \ \tau=p,n. \label{e3}
\end{equation}
Here $\rho^\tau_{\rm g.s.}(r)$ are the proton and neutron ground state (g.s.)
densities and $\delta_\tau$ the corresponding deformation lengths. Given an
appropriate choice of the g.s. proton and neutron densities it is natural to
represent the IS and IV parts of the total g.s. density as
\begin{equation}
 \rho^{0(1)}_{g.s.}(r)=\rho^n_{g.s.}(r)\pm\rho^p_{g.s.}(r).
 \label{e4}
\end{equation}
One can then generate, using the same BM prescription, the IS and IV parts of
the nuclear transition density by deforming (\ref{e4}) as
\begin{equation}
 \rho^{0(1)}_\lambda(r)=-\delta_{0(1)}\frac{d[\rho^n_{\rm g.s.}(r)\pm
 \rho^p_{\rm g.s.}(r)]}{dr}.
 \label{e5}
\end{equation}
The explicit expressions for the IS and IV deformation lengths are then easily
obtained, after some integration in parts, as
\begin{equation}
 \delta_{0}=\frac{N<r^{\lambda-1}>_n\delta_n+Z<r^{\lambda-1}>_p\delta_p}
{A<r^{\lambda-1}>_A},
 \label{e6a}
\end{equation}
\begin{equation}
 \delta_{1}=\frac{N<r^{\lambda-1}>_n\delta_n-Z<r^{\lambda-1}>_p\delta_p}
{N<r^{\lambda-1}>_n-Z<r^{\lambda-1}>_p}.
 \label{e6b}
\end{equation}
Here the radial momenta $<r^{\lambda-1}>_x\ (x=n,p,A)$ are obtained with the
neutron, proton and total g.s. densities as
\begin{equation}
 <r^{\lambda-1}>_x=\int_0^\infty\rho^x_{\rm g.s.}(r)r^{\lambda+1}dr\Big/
 \int_0^\infty\rho^x_{\rm g.s.}(r)r^2dr.
\end{equation}
The transition matrix element associated with a given component of nuclear
transition density ($y=n,p,0,1$) is
\begin{equation}
 M_y=\int_0^\infty \rho^y_{\lambda}(r)r^{\lambda+2}dr.
\label{e7}
\end{equation}
The ratios of the neutron/proton and IS/IV transition matrix elements are given
by
\begin{equation}
\frac{M_n}{M_p}=\frac{N<r^{\lambda-1}>_n\delta_n}{Z<r^{\lambda-1}>_p\delta_p},
 \label{e7a}
\end{equation}
\begin{equation}
 \frac{M_1}{M_0}=\frac{(N<r^{\lambda-1}>_n-Z<r^{\lambda-1}>_p)\delta_1}
 {(A<r^{\lambda-1}>_A)\delta_0}.
 \label{e7b}
\end{equation}
It is useful to note that there is a one-to-one correspondence between the
ratios of transition matrix elements in the two representations, and they are
related by
\begin{equation}
 M_n/M_p=(1+M_1/M_0)/(1-M_1/M_0).
 \label{e7c}
\end{equation}
If one assumes that the excitation is purely \emph{isoscalar} and the neutron
and proton densities have the same radial shape (scaled by the ratio $N/Z$) then
$\delta_n=\delta_p=\delta_0=\delta_1$,
\begin{equation}
 \frac{M_n}{M_p}=\frac{N}{Z}\ \ {\rm and}\
 \ \frac{M_1}{M_0}=\frac{N-Z}{A}=\varepsilon. \label{e8}
\end{equation}
Consequently, any significant deviation of the $M_n/M_p$ ratio from $N/Z$ (or
deviation of the $M_1/M_0$ ratio from $\varepsilon$) would directly indicate an
isovector mixing in the nuclear excitation.

Since the electric transition probabilities $B(E2$) for 2$^+_1$ states in
$^{18,20,22}$O isotopes have been measured, we can choose the proton deformation
length $\delta_p$ so that the experimental transition rate is reproduced by
$B_{\rm exp}(E2\uparrow)=e^2|M_p|^2$. As a result, the only free parameter to be
determined from the folding model analysis of the $(p,p')$ data is the neutron
deformation length $\delta_n$. All the transition matrix elements and other
deformation lengths can be directly obtained from $\delta_p$ and $\delta_n$
using Eqs.~(\ref{e3})-(\ref{e7b}). This feature is the main advantage of our
folding model approach compared to the standard DWBA or CC analysis using the
collective model prescription (\ref{e2}).

We note that the same $^{18,20,22}$O$(p,p')$ data have been studied in the
folding model using the microscopic nuclear transition densities calculated in
the quasiparticle random phase approximation (QRPA) \cite{El02}. In these
calculations \cite{El00,Be06,Gu06}, the QRPA proton transition density is scaled
to reproduce the experimental $B(E\lambda$) values, while the strength of the
neutron transition density is adjusted to the best DWBA or CC fit to the
$(p,p')$ data. The $M_n$ and $M_p$ transition matrix elements given by the
`scaled' QRPA transition densities are then compared with the empirical
estimates. Since different effective $NN$ interactions were used in the folding
calculations of Refs.~\cite{El00,Be06,Gu06}, it is of interest from the reaction
theory point of view to probe the microscopic QRPA transition densities in our
folding model analysis using the same effective $NN$ interaction. Therefore, in
addition to the BM transition densities (\ref{e3}), we have used in the present
work also the QRPA transition densities for 2$^+_1$ states in Oxygen isotopes
given by the continuum QRPA calculation by Khan {\it et al.} \cite{El02} to
calculate the inelastic FF. The proton and neutron g.s. densities obtained in
the Hartree-Fock-Bogoljubov study \cite{Gr01} were used in the folding model
calculation of the \pA optical potential.

\subsection{Folding model with complex CDM3Y6 interaction}
In our version \cite{Kh02} of the folding model, the central \nA potential is
evaluated in a Hartree-Fock manner as
\begin{equation}
  U=\sum_{j\in A}[<ij|v_{\rm D}|ij>+<ij|v_{\rm EX}|ji>], \label{p1}
\end{equation}
where $v_{\rm D(EX)}$ are the direct and exchange components of the effective
$NN$ interaction between the incident nucleon $i$ and nucleon $j$ bound in the
target $A$. The antisymmetrization gives rise to the exchange term in
Eq.~(\ref{p1}) which makes the \nA potential nonlocal in the coordinate space.
To separate the IS and IV contributions, one needs to make explicit the spin-
and isospin dependence of the (energy- and density dependent) $NN$ interaction
\begin{eqnarray}
v_{\rm D(EX)}(E,\rho,s)=v^{\rm D(EX)}_{00}(E,\rho,s)+
 v^{\rm D(EX)}_{10}(E,\rho,s)(\bm{\sigma\sigma}') \nonumber\\
  +  v^{\rm D(EX)}_{01}(E,\rho,s)(\bm{\tau\tau}')+
 v^{\rm D(EX)}_{11}(E,\rho,s)(\bm{\sigma\sigma}')(\bm{\tau\tau}'),
\label{p2}
\end{eqnarray}
where $s$ is the internucleon distance. The contribution from the spin dependent
terms ($v_{10}$ and $v_{11}$) in Eq.~(\ref{p2}) to the central \nA potential
(\ref{p1}) is exactly zero for the (spin-saturated) Oxygen targets considered in
the present work.

Using a realistic local approximation for the exchange term, the \nA potential
(\ref{p1}) can be obtained \cite{Kh02} in terms of the isoscalar ($U_{\rm IS}$)
and isovector ($U_{\rm IV}$) parts as
\begin{equation}
 U(E,\bm{R})=U_{\rm IS}(E,\bm{R})\pm U_{\rm IV}(E,\bm{R}),
\label{p3}
\end{equation}
where the + sign pertains to incident neutron and - sign to incident proton. The
second term in Eq.~(\ref{p3}) is the microscopic expression for the Lane
potential in Eq.~(\ref{e1}) as well as its prototype in Eq.~(\ref{e2}) for the
inelastic scattering FF. Each term in Eq.~(\ref{p3}) consists of the
corresponding direct and exchange potentials
\begin{eqnarray}
 U_{\rm IS}(E,\bm{R})=\int\{[\rho_n(\bm{r})+\rho_p(\bm{r})]
 v^{\rm D}_{00}(E,\rho,s) \nonumber\\
  +  [\rho_n(\bm{R},\bm{r})+\rho_p(\bm{R},\bm{r})]
 v^{\rm EX}_{00}(E,\rho,s)j_0(k(E,R)s)\}d^3r,
\label{p4}
\end{eqnarray}
\begin{eqnarray}
 U_{\rm IV}(E,\bm{R})=\int\{[\rho_n(\bm{r})-\rho_p(\bm{r})]
 v^{\rm D}_{01}(E,\rho,s) \nonumber\\
  +  [\rho_n(\bm{R},\bm{r})-\rho_p(\bm{R},\bm{r})]
 v^{\rm EX}_{01}(E,\rho,s)j_0(k(R)s)\}d^3r, \label{p5}
\end{eqnarray}
where $\rho(\bm{r},\bm{r}')$ is the (one-body) density matrix of the target
nucleus, with $\rho(\bm{r})\equiv\rho(\bm{r},\bm{r})$. $j_0(x)$ is the
zero-order spherical Bessel function and momentum $k(R)$ is determined from
\begin{equation}
 k^2(E,R)=\frac{2\mu}{{\hbar}^2}[E_{\rm c.m.}-{\rm Re}~U(R)-V_{\rm C}(R)].
\label{p6}
\end{equation}
Here, $\mu$ is the nucleon reduced mass, $U(R)$ and $V_{\rm C}(R)$ are the
nuclear and Coulomb parts of the OP, respectively. For a consistent description
of the elastic and inelastic \nA scattering, one needs to take into account
explicitly the multipole decomposition of the neutron and proton densities that
enter the folding calculation (\ref{p4})-(\ref{p5}). The details of the folding
calculation of $U_{\rm IS}$ and $U_{\rm IV}$ are the same as those given in
Ref.~\cite{Kh02}, excepting the use of a realistic local approximation for the
\emph{transition} density matrix taken from Ref.~\cite{Lo78}. We note that there
exists a more sophisticated version of the single-folding approach, known as the
g-folding model \cite{Am01}, where the nonlocal exchange potential is treated
exactly in the Schr\"odinger equation for the scattered wave, using the explicit
wave function for each single-particle state $|j>$ taken from the shell model.
In this sense, our approach is more flexible because one needs to use for the
folding input only the total proton and neutron densities $\rho_\tau(\bm{r})$
and the nuclear densities of any structure model can be used. In particular, the
use of the `collective model' prescription (\ref{e3}) for the transition
densities has allowed us to determine the IS and IV deformations of a nuclear
excitation.

For the effective interaction, we used the density- and isospin dependent CDM3Y6
interaction \cite{Kh97}. While the isoscalar density dependence of the CDM3Y6
interaction has been well tested in the folding model analyses of refractive \aA
and \AA scattering (see recent review in Ref.~\cite{Kh07}), its \emph{isovector}
density dependence was studied only recently in the CC analysis
\cite{Kh05,Kh07b} of the charge exchange \pn reaction exciting the 0$^+$
isobaric analog states of targets ranging from $^{6}$He to $^{208}$Pb. We recall
that the IS density dependence of the CDM3Y6 interaction was introduced
\cite{Kh97} as
\begin{eqnarray}
 v^{\rm D(EX)}_{00}(E,\rho,s)=F_{\rm IS}(E,\rho)v^{\rm D(EX)}_{00}(s),
\label{g1} \\
 F_{\rm IS}(E,\rho)=C_0[1+\alpha_0\exp(-\beta_0\rho)-\gamma_0\rho],
\label{g2}
\end{eqnarray}
where $v^{\rm D(EX)}_{00}(s)$ are the direct and exchange components of the
isoscalar M3Y-Paris interaction \cite{An83}. Parameters of $F_{\rm IS}$ were
chosen \cite{Kh97} to reproduce the saturation properties of symmetric nuclear
matter in the HF calculation. With a linear energy dependence included into
$C_0$, the IS interaction (\ref{g1}) reproduces very well the empirical energy
dependence of the IS nucleon OP in nuclear matter \cite{Kh95}.

For an accurate folding model analysis of the $^{18,20,22}$O$(p,p')$ data, it is
highly desirable to have a \emph{complex}, density- and isospin dependent $NN$
interaction for the input of the folding calculation (\ref{p3})-(\ref{p5}).
Following Ref.~\cite{Kh07b}, we have constructed in the present work, explicitly
for each energy, an imaginary IS density dependence of the same functional form
(\ref{g2}) and a complex IV density dependence of the M3Y-Paris interaction
\begin{eqnarray}
 v^{\rm D(EX)}_{01}(E,\rho,s)=F_{\rm IV}(E,\rho)v^{\rm D(EX)}_{01}(s),
\label{g3} \\
 F_{\rm IV}(E,\rho)=C_1[1+\alpha_1\exp(-\beta_1\rho)-\gamma_1\rho],
\label{g4}
\end{eqnarray}
where the parameters were adjusted to reproduce the JLM density- and isospin
dependent nucleon OP \cite{Je77} in the HF calculation of nuclear matter. All
radial shapes of $v^{\rm D(EX)}_{00(01)}(s)$ were kept unchanged as derived in
terms of three Yukawas from the M3Y-Paris interaction \cite{An83} (see the
explicit expressions for $v^{\rm D(EX)}_{00(01)}(s)$ in Ref.~\cite{Kh96}). The
isovector part of the folded \pA OP has been used in Ref.~\cite{Kh07b} as the FF
for the \pn reaction exciting the isobaric analog states, based on the isospin
coupling scheme. It turned out \cite{Kh07b} that the strength of the real
isovector interaction (\ref{g3}) is quite weak to account for the observed \pn
data and an enhancement of about 20-30\% is needed for a good CC description of
the \pn reaction. Therefore, we have scaled parameter $C_1$ of the real IV
density dependence (\ref{g4}) by a factor of 1.3 before using for the input of
the folding calculation (\ref{p3})-(\ref{p5}). The final parameters of the
complex density dependences $F_{\rm IS(IV)}(E,\rho)$ are presented in Table
\ref{t0}. We note that the central \pA potential (\ref{p3}) is supplemented by
the spin-orbital term obtained with the folding method of Ref.~\cite{Kh02} and
the spin-orbital terms of the CDM3Y6 interaction $v^{T=0,1}_{LS}(\rho,s)$ are
assumed to have the same IS and IV density dependences as those used for the
central terms.
\begin{table*}[htb]
\caption{Parameters of the complex IS and IV density dependence of the CDM3Y6
interaction defined in Eqs.~(\ref{g2}) and (\ref{g4}), respectively.}\label{t0}
\begin{ruledtabular}
\begin{tabular}{c|c|c|c|c|c|c|c|c} 
 \multicolumn{5}{c|}{Re~$F_{\rm IS}(E,\rho)$} &
 \multicolumn{4}{c}{Im~$F_{\rm IS}(E,\rho)$}\\ \hline
$E$ (MeV) & $C_0$ & $\alpha_0$ & $\beta_0$ (fm$^3$) & $\gamma_0$ (fm$^3$) &
   $C_0$ & $\alpha_0$ & $\beta_0$ (fm$^3$) & $\gamma_0$ (fm$^3$)\\ \hline
24.5 & 0.2487 & 3.8033 & 1.4099 &  4.0   & 0.1504 & 6.0964 & 15.503 & 4.3931 \\
43.0 & 0.2361 & 3.8033 & 1.4099 &  4.0   & 0.0869 & 8.9092 &  9.9237 & 4.2128 \\
46.6 & 0.2336 & 3.8033 & 1.4099 &  4.0   & 0.1029 & 6.8937 &  9.1076 & 4.2056 \\
 \hline
 \multicolumn{5}{c|}{Re~$F_{\rm IV}(E,\rho)$} &
 \multicolumn{4}{c}{Im~$F_{\rm IV}(E,\rho)$}\\ \hline
$E$ (MeV) & $C_1$ & $\alpha_1$ & $\beta_1$ (fm$^3$) & $\gamma_1$ (fm$^3$) &
  $C_1$ & $\alpha_1$ & $\beta_1$ (fm$^3$) & $\gamma_1$ (fm$^3$)\\ \hline
24.5 & 0.2668 & 6.3227 & 13.725 & -3.8888 & 0.2010 & 9.6207 & 16.053 & -4.3670 \\
43.0 & 0.1490 & 9.7964 & 10.743 & -4.1147 & 0.2315 & 6.2846 & 13.162 & -4.3612 \\
46.6 & 0.1585 & 9.3490 & 11.683 & -3.9323 & 0.2289 & 6.0590 & 12.407 & -4.4283 \\
\end{tabular}
\end{ruledtabular}
\end{table*}
All the optical model (OM) and CC calculations have been performed using the CC
code ECIS97 written by Raynal \cite{Ra97}.

\section{Results and discussions}
 \label{sec3}
\subsection{Transition strength of 2$^+_1$ state in $^{18}$O}
For the Oxygen isotopes under study, the 2$^+_1$ state (at 1.98 MeV) in $^{18}$O
is the most studied one. The core polarization by the two valence neutrons in
the 2$^+_1$ excitation of $^{18}$O was shown to be quite strong, with the
$M_n/M_p$ ratio significantly larger than $N/Z$. The neutron transition strength
of 2$^+_1$ state in $^{18}$O has been measured in several experiments, like the
(direct) inelastic proton and neutron scattering \cite{Gr80,Es74,Ke86} or
inelastic pion scattering \cite{Se88}. The $M_n/M_p$ ratio was often deduced by
the (collective model) prescription of Bernstein, Brown and Madsen (BBM)
\cite{Be83} which has been checked against the data collected for a wide range
of single-closed shell nuclei. While the BBM analysis of the low-energy proton
scattering data seems to favor $M_n/M_p\approx 1.5$ for 2$^+_1$ state in
$^{18}$O \cite{Je99,Ct01}, the inelastic pion scattering data were shown to give
a much higher $M_n/M_p$ ratio of 2.3 to 2.4 \cite{Se88}. The $M_n/M_p$ ratio
deduced from the pion scattering data also agrees fairly with that deduced from
the measured $B(E2)$ strength of 2$^+_1$ state in the mirror nucleus $^{18}$Ne,
using the isospin symmetry \cite{Be79}. We note that the DWBA analysis of
$(p,p')$ and $(n,n')$ scattering data at 24 MeV \cite{Gr80} is of particular
interest for our study, as it is the only attempt to determine the IS and IV
deformations for 2$^+_1$ state in $^{18}$O prior to our work \cite{Kh03}. By
using prescription (\ref{e2}) and assuming $\delta_0$ to be the average of
$\delta$ values given by the $(p,p')$ and $(n,n')$ data, the authors of
Ref.~\cite{Gr80} obtained $\delta_0\approx 1.26\pm 0.06$ fm and $\delta_1\approx
3.14\pm 1.57$ fm which correspond to $M_n/M_p\approx 1.72\pm 0.70$. The
coupled-channel effect at the proton energy of 24 MeV was shown \cite{Gr80} to
affect slightly the deduced deformation parameters.

To compare our microscopic folding model analysis of $^{18}$O$(p,p')$ data with
the collective model results of Ref.~\cite{Gr80}, we have performed a
coupled-channel ($2^+_1\leftrightarrow 0^+_{\rm g.s.}\leftrightarrow 3^-_1$)
analysis of $^{18}$O$(p,p')$ data at 24.5 MeV \cite{Es74} using the (complex) OP
and inelastic FF given by the folding calculation (\ref{p3})-(\ref{p5}) for the
lowest 2$^+$ and 3$^-$ states in $^{18}$O. By adjusting $M_p$ to the
experimental $B(E2\uparrow)=45.1\pm 2.0\ e^2$fm$^4$ \cite{Ra01} and
$B(E3\uparrow)=1120\pm 11\ e^2$fm$^6$ \cite{Sp89} for the first 2$^+$ and 3$^-$
states in $^{18}$O, we obtain $\delta_p=1.04\pm 0.02$ and $1.45\pm 0.01$ fm,
respectively, for the corresponding proton transition densities (\ref{e3}). To
effectively account for the higher-order dynamic polarization of the OP by the
open nonelastic channels, the (complex) strength of the CDM3Y6 interaction is
first adjusted to the best CC description of elastic scattering data and then is
used without any further renormalization to calculate the inelastic FF. As a
result, the only remaining parameter is the neutron deformation length
$\delta_n$ which is determined from the best CC fit to the inelastic scattering
data.
\begin{figure}[htb]
\mbox{\epsfig{file=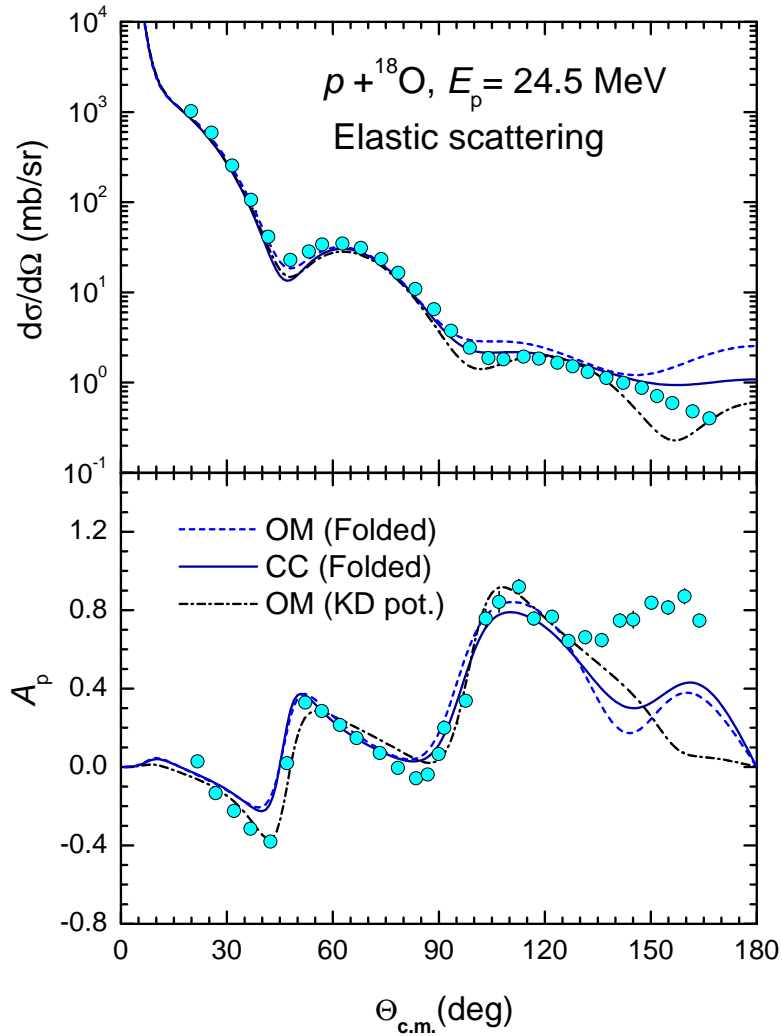,height=15cm}} \vspace*{-1cm}
 \caption{(Color online) Measured differential cross section and analyzing
power of the elastic $p+^{18}$O scattering at 24.5 MeV \cite{Es74} versus the OM
results given by the microscopic folded OP and phenomenological OP taken from
the global systematics \cite{Ko03} by Koning and Delaroche (KD). The CC results
are given by the complex OP and inelastic FF calculated in the folding model.}
\label{f1}
\end{figure}
The renormalization factors $N_{\rm R}$ and $N_{\rm I}$  of the real and
imaginary folded OP were first obtained in the OM analysis of the elastic data
(see Table~\ref{t1}). At 24.5 MeV, the OM fit gives $N_{\rm R}\approx 0.91$ and
$N_{\rm I}\approx 0.6$. These values have changed slightly to $N_{\rm R}\approx
0.97$ and $N_{\rm I}\approx 0.57$ when the two-channel coupling is taken into
account. The folded spin-orbital potential needs a renormalization of around 0.5
in both the OM and CC calculations. At higher energy of 43 MeV the best-fit
$N_{\rm R}$ factor becomes close to unity while $N_{\rm I}$ remains around 0.6.
One can see from Fig.~\ref{f1} that the folded OP gives quite a good description
of the measured elastic cross section and analyzing power at 24.5 MeV in both
the OM and CC schemes.
\begin{table*}[htb]
\caption{Renormalization factors $N_{\rm R}$ and $N_{\rm I}$ of the real and
imaginary folded potentials used in the OM and CC calculations of elastic and
inelastic $p+^{18,20,22}$O scattering.} \label{t1}
\begin{ruledtabular}
\begin{tabular}{c|c|c|c|c|c|c|c|c}
  \multicolumn{6}{c|}{OM fit} &
 \multicolumn{3}{c}{CC fit}\\ \hline
Target & $E$ (MeV) & $N_{\rm R}$ & $N_{\rm I}$ & $N_{\rm LS}$ &
 $\sigma_{\rm R}$ (mb) & $N_{\rm R}$ & $N_{\rm I}$ & $N_{\rm LS}$ \\ \hline
 $^{18}$O & 24.5 & 0.91 & 0.60 & 0.50 & 658 & 0.97 & 0.57 & 0.50 \\
 $^{18}$O & 43.0 & 1.00 & 0.69 & 0.50 & 545 & 1.05 & 0.64 & 0.50 \\
 $^{20}$O & 43.0 & 1.08 & 0.68 & 0.50 & 573 & 1.15 & 0.60 & 0.50 \\
 $^{22}$O & 46.6 & 1.00 & 0.72 & 0.50 & 572 & 1.01 & 0.70 & 0.50 \\
\end{tabular}
\end{ruledtabular}
\end{table*}
For a comparison, we have also performed the OM calculation using the
phenomenological OP (parameterized in terms of Woods-Saxon potentials) taken
from an accurate global systematics \cite{Ko03} by Koning and Delaroche (KD).
Although KD systematics has been developed for nuclei in the mass range
$24\leqslant A\leqslant 209$, our OM analysis shows that it works rather well
also for the Oxygen isotopes under study.

\begin{figure}[htb]
\mbox{\epsfig{file=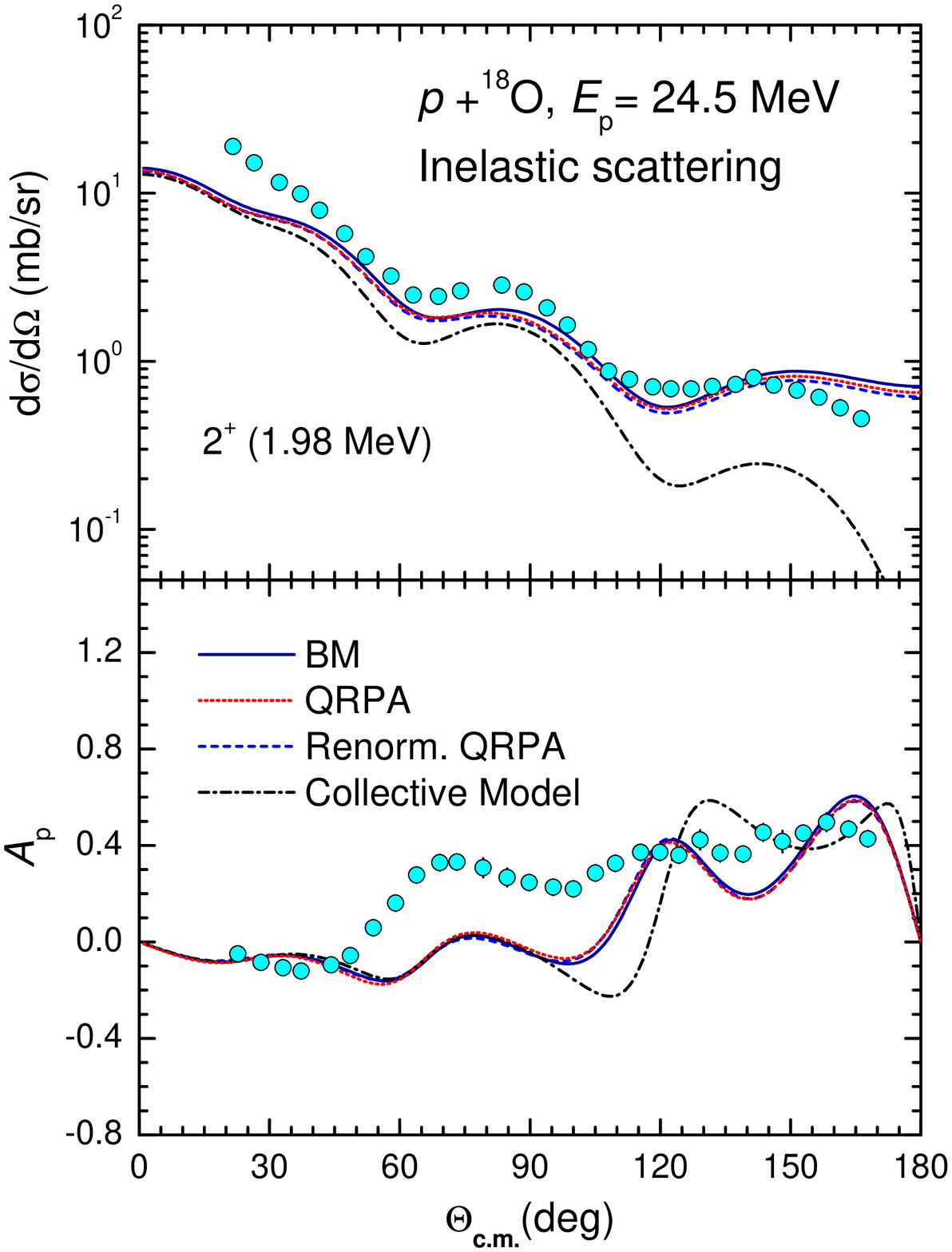,height=15cm}} \vspace*{-1cm}
 \caption{(Color online) Measured differential cross section and analyzing
power of the inelastic $p+^{18}$O scattering at 24.5 MeV \cite{Es74} versus the
CC results given by the microscopic folded FF obtained with BM and QRPA
transition densities. The collective model result is given by form factor
(\ref{e2}) obtained with the phenomenological OP by Koning and Delaroche
\cite{Ko03} (see more details in text).} \label{f2}
\end{figure}
The inelastic $p+^{18}$O scattering data at 24.5 MeV \cite{Es74} for 2$^+_1$
state in $^{18}$O are compared with the results of CC calculation in
Fig.~\ref{f2}. Like the earlier folding model study \cite{Kh03}, we found in the
present analysis a significant IV mixing in 2$^+_1$ excitation which leads to
$M_n/M_p\approx 1.55$ (see Table~\ref{t2}). This value is about 25\% larger than
the ratio implied by the isoscalar limit ($M_n/M_p=N/Z=1.25$). Compared to the
elastic channel, the agreement of the CC results with the measured analyzing
power of inelastic scattering is rather poor at medium angles, and that could
well be due to a simple treatment the inelastic spin-orbital FF adopted in our
folding method \cite{Kh02}. However, the inelastic spin-orbital FF does not
affect significantly the calculated inelastic 2$^+$ scattering cross section
which is dominated by contribution from the central FF, and the widely accepted
procedure is to deduce deformation parameters by matching the calculated
inelastic scattering cross section to the data. To stress the reliability of the
folding approach, we have done in parallel the same CC calculation but using the
collective model form factor (\ref{e2}) determined with the phenomenological OP
by Koning and Delaroche \cite{Ko03} and the same IS and IV deformation lengths.
As expected, the form factor (\ref{e2}) was found to strongly underestimate the
measured 2$^+$ cross section at large angles, in about the same way as
established earlier in the folding model studies of inelastic heavy-ion
scattering \cite{Be95,Kh00}. This explains naturally why the IS and IV
deformation lengths of 2$^+_1$ state in $^{18}$O deduced by Grabmayr {\it et
al.} \cite{Gr80} in their collective model analysis of the same data
($\delta_0\approx 1.26$ fm and $\delta_1\approx 3.14$ fm) are significantly
larger than the values deduced from our folding model analysis (see
Table~\ref{t2}).
\begin{figure}[htb]
\mbox{\epsfig{file=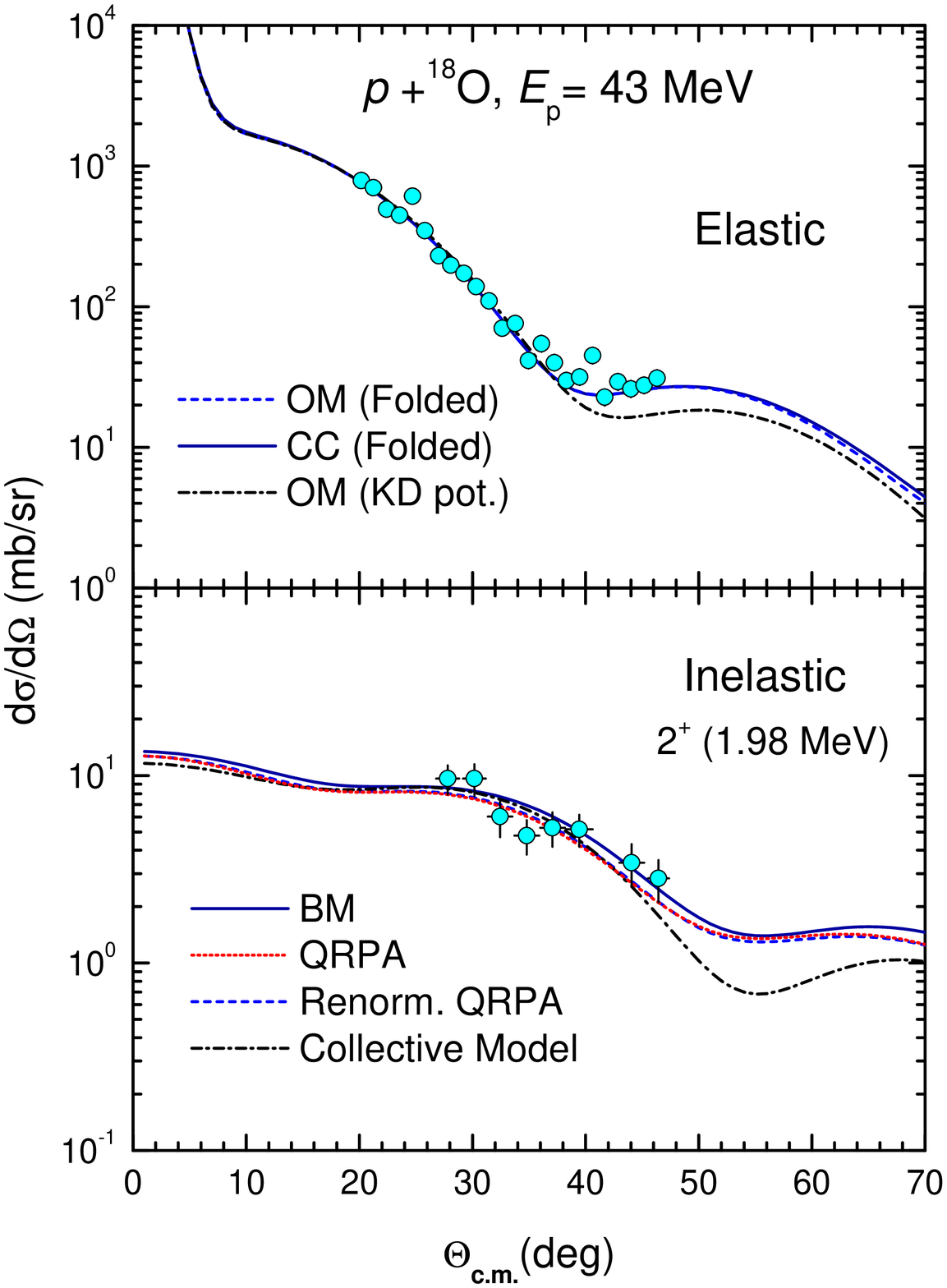,height=15cm}} \vspace*{-1cm}
 \caption{(Color online) Elastic and inelastic $p+^{18}$O scattering data
at 43 MeV \cite{El00} in comparison with the OM and CC results. Notations for
the OP and inelastic FF are the same as used in Figs.~\ref{f1} and \ref{f2}.}
\label{f3}
\end{figure}
The numerical uncertainties of $\delta_{p}$ given in Table~\ref{t2} are fully
determined by those of the measured $B(E2)$ values, while an uncertainty of
around 5\% was assigned to $\delta_{n}$ which gives a cross-section shift within
the experimental errors. The uncertainties of the IS and IV deformation lengths
and ratios of transition matrix elements were deduced directly from those found
for $\delta_p$ and $\delta_n$.

The neutron deformation length found in the CC analysis of the inelastic
$p+^{18}$O scattering data at 24.5 MeV has been used to calculate the inelastic
$p+^{18}$O scattering FF at higher energy of 43 MeV. With only strength of the
complex CDM3Y6 interaction slightly adjusted by the CC fit to elastic scattering
data, the folding + CC description of the measured $(p,p')$ data at 43 MeV
\cite{El00} is quite satisfactory (see Fig.~\ref{f3}) without any further
adjustment of $\delta_{n}$ for 2$^+_1$ state in $^{18}$O. Given the proton
transition strength fixed by the measured $B(E2)$ value, we conclude that the
neutron transition matrix element $M_n$ should be such that the ratio
$M_n/M_p\approx 1.55$ or equivalently $M_1/M_0\approx 0.22$. The latter is two
times larger than $\varepsilon=0.11$ and, hence, implies a significant IV mixing
in the 2$^+_1$ excitation. Our present result also agrees closely with that
given by the BBM analysis of low-energy proton scattering data ($M_n/M_p\approx
1.5$ for 2$^+_1$ state in $^{18}$O) \cite{Je99,Ct01}.
\begin{table*}[htb]
\caption{Deformation lengths and ratios of the transition matrix elements for
2$^+_1$ states in $^{18,20,22}$O deduced from the present folding + CC analysis
of inelastic proton scattering [see definitions in
Eqs.~(\ref{e3})-(\ref{e7b})].} \label{t2}
\begin{ruledtabular}
\begin{tabular}{|c|c|c|c|c|c|c|c|c|}
 Nucleus & $N/Z$ & $\varepsilon$ & $\delta_p$ (fm) & $\delta_n$ (fm) &
 $M_n/M_p$ & $\delta_0$ (fm) & $\delta_1$ (fm) & $M_1/M_0$ \\ \hline
 $^{18}$O & 1.25 & 0.11 & $1.04\pm 0.02$ & $1.23\pm 0.06$ &
 $1.55\pm 0.08$ & $1.15\pm 0.04$ & $1.87\pm 0.27$ & $0.22\pm 0.03$
 \\ \hline
 $^{20}$O & 1.50 & 0.20 & $0.82\pm 0.03$ & $1.95\pm 0.10$ &
 $3.24\pm 0.20$ & $1.52\pm 0.06$ & $3.75\pm 0.26$ & $0.59\pm 0.05$
 \\ \hline
 $^{22}$O & 1.75 & 0.27 & $0.70\pm 0.12$ & $0.91\pm 0.05$ &
 $1.81\pm 0.33$ & $0.85\pm 0.05$ & $1.14\pm 0.16$ & $0.43\pm 0.07$
 \\ \end{tabular}
\end{ruledtabular}
\end{table*}
The microscopic QRPA transition densities \cite{El02} also give a satisfactory
description of the inelastic $p+^{18}$O scattering data under study (see
Figs.~\ref{f2} and \ref{f3}). We note that the continuum QRPA calculation of
Ref.~\cite{El02} strongly underestimates the $E2$ strength and gives
$B(E2\uparrow)=14\ e^2$fm$^4$ for 2$^+_1$ state in $^{18}$O compared to the
experimental value of 45 $e^2$fm$^4$. On the contrary, the predicted neutron
transition strength is much too high ($M_n/M_p\approx 2.88$) compared to that
found in the present work and other studies. Such a strong neutron transition
strength seems to compensate for the weak proton transition strength predicted
by the QRPA, and the inelastic FF folded with the QRPA transition densities
gives a reasonable description of the inelastic scattering data. It is
interesting that about the same good description of the inelastic scattering
data is given by the \emph{renormalized} QRPA densities, with $\rho^p_2(r)$
scaled to reproduce the measured $B(E2)$ value and $\rho^n_2(r)$ scaled to give
the same $M_n$ as that obtained above for the BM model (\ref{f3}) of transition
densities (see Fig.~\ref{f2} and lower panel of Fig.~\ref{f3}).

\begin{figure}[htb]
\mbox{\epsfig{file=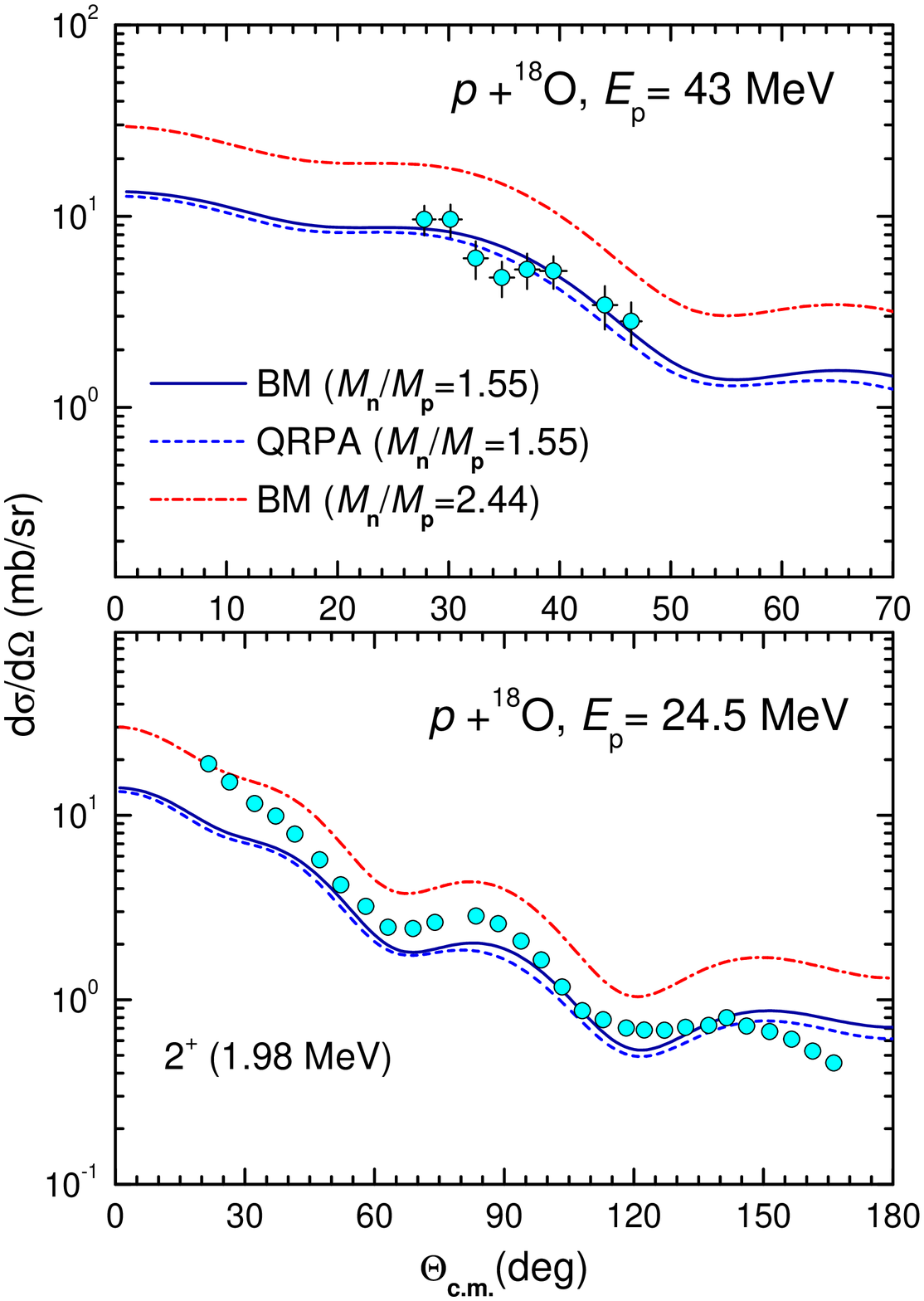,height=15cm}} \vspace*{-0.5cm}
 \caption{(Color online) Inelastic $p+^{18}$O scattering data at 24.5 \cite{Es74}
and 43 MeV \cite{El00} for 2$^+_1$ state in $^{18}$O in comparison with the CC
results. The neutron deformation length $\delta_n$ of the transition density
(\ref{e3}) was adjusted to give two different ratios $M_n/M_p=1.55$ and 2.44.}
\label{f4}
\end{figure}
As mentioned above, an alternative method to determine the neutron transition
matrix element $M_n$ has been suggested some 30 years ago by Bernstein {\it et
al.} \cite{Be79} based on the isospin symmetry. Namely, $M_n$ can be obtained
from $M_p$ measured for the same $2^+$ excitation in the mirror nucleus with an
electromagnetic probe if one assumes the charge independence of the $2^+$
excitation in members of a $T$-isospin multiplet. In particular, one has for the
isobars with opposite signs of the isospin projection $T_z$
\begin{equation}
 M_p(-T_z)=M_n(T_z). \label{is1}
\end{equation}
Using the electric transition rates $B(E2)$ for $^{18}$O and $^{18}$Ne taken
from the latest compilation of the experimental data \cite{Ra01}, we easily
deduce the ratio of transition matrix elements $M_n/M_p\approx 2.44\pm 0.18$ for
$2^+_1$ states in $^{18}$O using Eq.~(\ref{is1}). This value is significantly
larger than that obtained in the present folding + CC analysis and collective
model analyses reported in Refs.~\cite{Je99,Ct01,Gr80}. To illustrate such a
difference in terms of the $(p,p')$ cross section, we have done the same folding
+ CC calculation but using the \emph{enhanced} neutron transition density
(\ref{e3}) which gives $M_n/M_p\approx 2.44$. Then, the calculated inelastic
cross sections strongly overestimate the measured $(p,p')$ data at both energies
under study (see Fig.~\ref{f4}). Assuming the realistic value $M_n/M_p\approx
1.55$, we might interpret the difference shown in Fig.~\ref{f4} as an indication
to the isospin impurity in the $2^+_1$ excitations of the $A=18,\ T=1$ isobaric
multiplet. It is complementary to note that a similar isospin-impurity effect
has been found by Khan {\it et al.} \cite{El04} for the $2^+_1$ excitations of
the $A=30,\ T=1$ isobaric multiplet. Nevertheless, if one takes into account the
data of inelastic $\pi^+$ and $\pi^-$ scattering from $^{18}$O \cite{Se88} which
give $M_n/M_p\approx 2.3-2.4$ in a distorted-wave impulse approximation analysis
using different types of the transition density for $2^+_1$ state, then the
measured $B(E2)$ value of $2^+_1$ state in the mirror $^{18}$Ne nucleus seems to
support a good isospin symmetry in this case. Given the accurately measured
electric transition rates $B(E2)$ of $2^+_1$ states in $^{18}$O and $^{18}$Ne, a
future (high-precision) experiment to re-determine the neutron transition
strengths of $2^+_1$ states in these two mirror nuclei should provide vital data
for the determination of the isospin impurity using relation (\ref{is1}).

\subsection{Transition strength of 2$^+_1$ state in $^{20}$O}
In difference from the stable $^{18}$O target, inelastic proton scattering to
2$^+_1$ state in the \emph{unstable} $^{20}$O isotope has been measured only
recently, in the inverse kinematics, at 30 MeV \cite{Je99} and 43 MeV
\cite{El00}. Using the measured transition rate $B(E2\uparrow)\approx 28.1\pm
2.0\ e^2$fm$^4$ \cite{Ra01} for 2$^+_1$ state in $^{20}$O, we have deduced the
proton deformation length $\delta_p\approx 0.82\pm 0.03$ fm for the proton
transition density (\ref{e3}). The proton deformation length of 3$^-_1$ state in
$^{20}$O was taken from the empirical estimate of Ref.~\cite{Kh03}, and the
neutron deformation length was then adjusted to the best CC fit to the inelastic
$^{18}$O$(p,p')$ data for 3$^-_1$ excitation \cite{El00}. Similar to our earlier
folding + DWBA analysis \cite{Kh03} of these data, the best-fit $M_n/M_p$ ratio
for the 3$^-_1$ excitation turned out to be quite close to $N/Z=1.5$ which
indicates a dominant IS character of this state.
\begin{figure}[htb]
\mbox{\epsfig{file=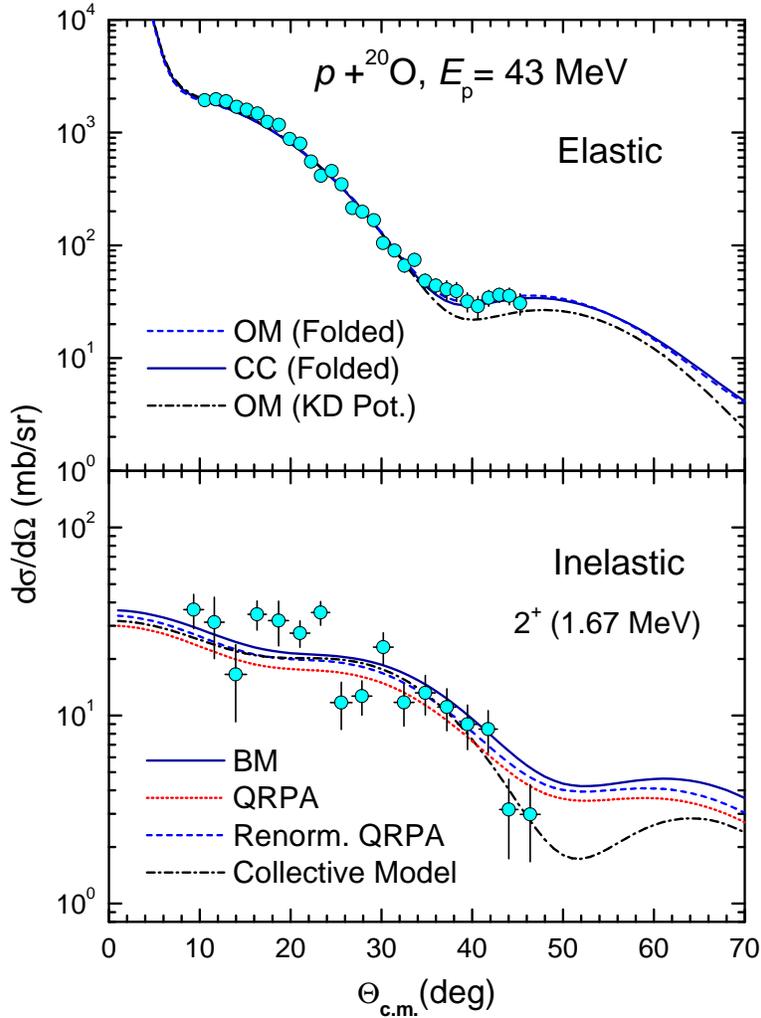,height=15cm}} \vspace*{-0.5cm}
 \caption{(Color online) Elastic and inelastic $p+^{20}$O scattering data at
43 MeV \cite{El00} in comparison with the OM and CC results. Notations for the
OP and inelastic FF are the same as used in Figs.~\ref{f1} and \ref{f2}.}
\label{f5}
\end{figure}
The isospin character of 2$^+_1$ state is very much different from that found
for 3$^-_1$ state. Using the deformation parameters extracted from the CC
analysis of the 30 MeV data with the collective model FF, the ratio of
neutron/proton transition matrix elements $M_n/M_p\approx 2.9\pm 0.4$  has been
deduced by Jewell {\it et al.} \cite{Je99} for 2$^+_1$ state in $^{20}$O. The
JLM folding model analysis of 43 MeV data by Khan {\it et al.} \cite{El00} using
the QRPA transition densities has found a stronger IV mixing in this state, with
$M_n/M_p\approx 3.25\pm 0.80$. Our earlier folding model analysis \cite{Kh03},
using the compact method (\ref{e3})-(\ref{e7b}) and original CDM3Y6 interaction
with the IV density dependence assumed to be the same as the IS one, has given a
larger ratio of $M_n/M_p\simeq 4.2$ for 2$^+_1$ states in $^{20}$O. In the
present work we concentrate on the 43 MeV data which contain more data points
and cover a wider angular range. The results of our folding + CC analysis are
compared with the elastic and inelastic $p+^{20}$O scattering data at 43 MeV in
Fig.~\ref{f5}. With $\delta_p$ fixed above by the measured $B(E2)$ value, the
best-fit neutron deformation length is $\delta_n\approx 1.95$ fm which results
on the ratios $M_n/M_p\approx 3.24$ and $M_1/M_0\approx 0.59$. These values are
well exceeding the IS limit of $M_n/M_p\approx N/Z=1.5$ and $M_1/M_0\approx
\varepsilon=0.2$. The deduced IV deformation length (see Table~\ref{t2}) is
about 2.5 times the IS deformation length and confirms, therefore, a strong core
polarization by the valence neutrons in 2$^+_1$ excitation of the
\emph{open-shell} $^{20}$O nucleus. In difference from the inelastic $p+^{18}$O
scattering data at 24.5 MeV shown in Fig.~\ref{f2}, the 43 MeV data for
$^{18,20}$O `targets' (measured at angles $\Theta_{\rm c.m.} < 50^\circ$ only)
are reasonably reproduced by both the folded and collective model form factors
based on the same $\delta_0$ and $\delta_1$. A substantial difference between
the CC results given by these two choices of inelastic FF was found at larger
scattering angles ($50^\circ < \Theta_{\rm c.m.} < 180^\circ$) where no data
point was taken. In this sense, a future experiment aiming to measure $(p,p')$
cross section over a wider angular range could provide a better test ground for
the inelastic FF and neutron transition strength.

The continuum QRPA description of 2$^+_1$ state in $^{20}$O is better than that
for $^{18}$O, with the predicted $B(E2\uparrow)\approx 22\ e^2$fm$^4$ (compared
to the adopted experimental value of 28 $e^2$fm$^4$) and $M_n/M_p\approx 3.36$
\cite{El02}. A good CC description of the $(p,p')$ cross section was obtained
after a slight renormalization of the QRPA transition densities to reproduce the
experimental $B(E2)$ value and best-fit $M_n/M_p$ ratio (see Fig.~\ref{f5}). We
stress that the use of a more realistic version of the (complex) density- and
isospin dependent CDM3Y6 interaction in the present work has pinned down the
best-fit ratio of transition matrix elements for 2$^+_1$ state in $^{20}$O to
$M_n/M_p\approx 3.24$ which is very close to that deduced from the JLM folding
model analysis \cite{El00} of the same data. Together with the results of the
BBM analysis reported in Refs.~\cite{Je99,Ct01}, our results confirm again a
strong IV mixing in 2$^+_1$ excitation of $^{20}$O.

Like the $^{18}$O case, there is an alternative method to determine the
$M_n/M_p$ ratio from the electric $B(E2)$ transition rates measured for 2$^+_1$
states in $^{20}$O and its mirror partner $^{20}$Mg, using Eq.~(\ref{is1}) given
by the isospin symmetry. Towards this goal, a measurement of the Coulomb
excitation of the unstable $^{20}$Mg nucleus has been performed by RIKEN group
\cite{Iw08} using a radioactive $^{20}$Mg beam incident on the lead target. The
extracted transition rate $B(E2\uparrow)\approx 177\pm 32\ e^2$fm$^4$ for
2$^+_1$ state in $^{20}$Mg seems to agree well with the prediction of realistic
cluster model for this nucleus. Assuming the proton transition matrix element
$M_p$ for 2$^+_1$ state in $^{20}$Mg equal the neutron transition matrix element
$M_n$ for 2$^+_1$ state in $^{20}$O, one obtains easily $M_n/M_p\approx 2.51\pm
0.25$ for the latter. This value is about 30\% smaller than the best-fit
$M_n/M_p$ ratio obtained in Ref.~\cite{El00} and present work. Given an accurate
treatment of the folding model ingredients and similar effect discussed above
for 2$^+_1$ state in $^{18}$O, such a difference in $M_n/M_p$ ratios deduced by
the two methods might well indicate the isospin impurity in the $2^+_1$
excitation of the $A=20,\ T=2$ isobaric multiplet.
\begin{figure}[htb]
 \vspace*{-1cm}
\mbox{\epsfig{file=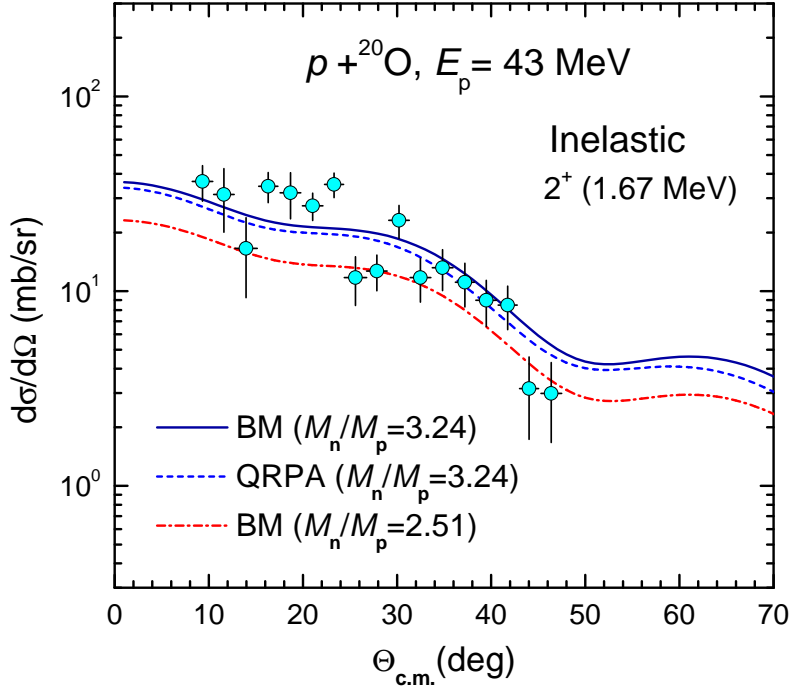,width=12cm}} \vspace*{-4cm}
 \caption{(Color online) Inelastic $p+^{20}$O scattering data at 43 MeV
\cite{El00} for 2$^+_1$ state in $^{20}$O in comparison with the CC results. The
neutron deformation length $\delta_n$ of the transition density (\ref{e3}) was
adjusted to give two different ratios $M_n/M_p=2.51$ and 3.24.} \label{f6}
\end{figure}
To illustrate this effect in the calculated inelastic cross sections, we have
plotted in Fig.~\ref{f6} the CC result obtained with two choices of the neutron
transition density (\ref{e3}) which were \emph{scaled} to give $M_n/M_p\approx
2.51$ and 3.24. One can see that the CC results associated with $M_n/M_p\approx
2.51$ substantially underestimate the measured $(p,p')$ data. In terms of the
total $(p,p')$ cross section, the difference caused by the `isospin impurity'
shown in Fig.~\ref{f6} is around 40\%. This difference reduces to around 30\%
when one adopts the upper limit of the measured transition rate,
$B(E2\uparrow)\approx 210\ e^2$fm$^4$, for 2$^+_1$ state in $^{20}$Mg
\cite{Iw08}. We must note, however, that the last 2 data points in Fig.~\ref{f6}
seem to agree better with the CC results associated with $M_n/M_p\approx 2.51$.
In the same logic as discussed above for $^{18}$O and $^{18}$Ne, a future
\emph{inverse-kinematics} measurement of $^{20}$O$(p,p')$ and $^{20}$Mg$(p,p')$
reactions to determine neutron transition strength of 2$^+_1$ states in these
two mirror \emph{unstable} nuclei could provide important data for the check of
isospin purity in the $2^+_1$ excitation of the $A=20,\ T=2$ isobars using
relation (\ref{is1}).

\subsection{Transition strength of 2$^+_1$ state in $^{22}$O}
If we consider $^{20}$O as consisting of the $^{16}$O core and four valence
neutrons, then the large IV deformation length $\delta_1$ extracted above for
2$^+_1$ state in $^{20}$O indicates a strong core polarization by the valence
neutrons in the 2$^+_1$ excitation. In such a `core + valence neutrons' picture,
it is natural to expect that 2$^+_1$ state in $^{22}$O should be more collective
and have a larger IV deformation length due to the contribution of two more
valence neutrons. However, the inelastic $^{22}$O$(p,p')$ scattering data at
46.6 MeV measured recently at GANIL \cite{Be06} show clearly the opposite
effect, with the inelastic cross section about 3 to 4 times smaller than
$^{22}$O$(p,p')$ cross section measured  at 43 MeV \cite{El00} for $2^+_1$ state
in $^{20}$O. The folding + DWBA analysis \cite{Be06} of these data using the
method (\ref{e3})-(\ref{e7b}) and original CDM3Y6 interaction (with the IV
density dependence assumed to be the same as the IS one) has pointed to a much
weaker neutron transition strength of $2^+_1$ in $^{22}$O. Given a significantly
higher excitation energy of this state (1.5 MeV higher than that of $2^+_1$
state in $^{20}$O), the $^{22}$O$(p,p')$ data at 46.6 MeV were considered
\cite{Be06} as an important evidence for the neutron shell closure occurring at
$N=14$ or 16. Given the measured transition rate $B(E2\uparrow)\approx 21\pm 8\
e^2$fm$^4$ \cite{Ra01} for 2$^+_1$ state in $^{22}$O, we have obtained the
proton deformation length $\delta_p\approx 0.70\pm 0.12$ fm for the proton
transition density (\ref{e3}). Like in cases of $^{18,20}$O, the neutron
deformation length for $2^+_1$ state in $^{20}$O was adjusted to the best
description of the 46.6 MeV data by the folding + CC analysis. Since there are
no data measured for $(p,p')$ scattering at 46.6 MeV to 3$^-_1$ state in
$^{22}$O, we have considered in our CC scheme only the coupling between the
elastic and 2$^+$ inelastic scattering channels ($0^+_{\rm g.s.}\leftrightarrow
2^+_1$).
\begin{figure}[htb]
\mbox{\epsfig{file=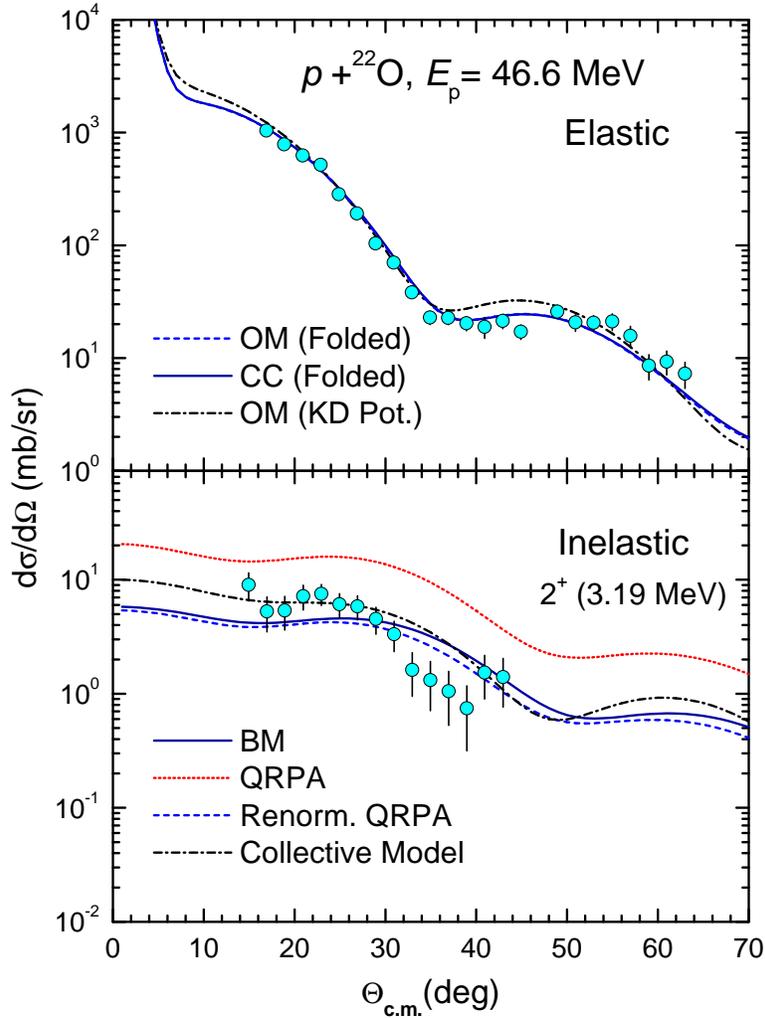,height=15cm}} \vspace*{-0.5cm}
 \caption{(Color online) Elastic and inelastic $p+^{22}$O scattering data at
46.6 MeV \cite{Be06} in comparison with the OM and CC results. Notations for the
OP and inelastic FF are the same as used in Figs.~\ref{f1} and \ref{f2}.}
\label{f7}
\end{figure}
The folding + CC results are compared with the elastic and inelastic $p+^{22}$O
scattering data in Fig.~\ref{f7}. With $\delta_p$ fixed by the measured $B(E2)$
value, the best-fit neutron deformation length was found $\delta_n\approx 0.9$
fm which lead to the ratios $M_n/M_p\approx 1.81$ and $M_1/M_0\approx 0.43$.
These values are rather close to those implied by the IS limit of
$M_n/M_p\approx N/Z=1.75$ and $M_1/M_0\approx \varepsilon=0.27$. The deduced IV
deformation length $\delta_1$ is around 30\% larger than the IS deformation
length $\delta_0$ (see Table~\ref{t2}) and this is much smaller than the
difference between $\delta_1$ and $\delta_0$ found above for 2$^+_1$ state in
$^{20}$O. Thus, our results show a much weaker polarization effect by the
valence neutrons in the 2$^+_1$ excitation of $^{22}$O nucleus. This subtle
effect could not be accurately described by the continuum QRPA. While the QRPA
calculation gives $B(E2\uparrow)\approx 22\ e^2$fm$^4$ (in a perfect agreement
with the measured value \cite{Ra01}), the predicted $M_n/M_p\approx 3.53$
\cite{El02} is nearly two times the empirical data. That is the reason why the
CC results given by the original QRPA transition densities for 2$^+_1$ state in
$^{22}$O strongly overestimate the data as shown in Fig.~\ref{f7}. Like the
$^{20}$O case, the inelastic $p+^{22}$O data are reasonably reproduced by both
the folded and collective model FF based on the same deformation IS and IV
lengths. A significant difference given by the two choices of inelastic FF can
be seen at larger scattering angles where no data point was measured.

\begin{figure}[htb]
\mbox{\epsfig{file=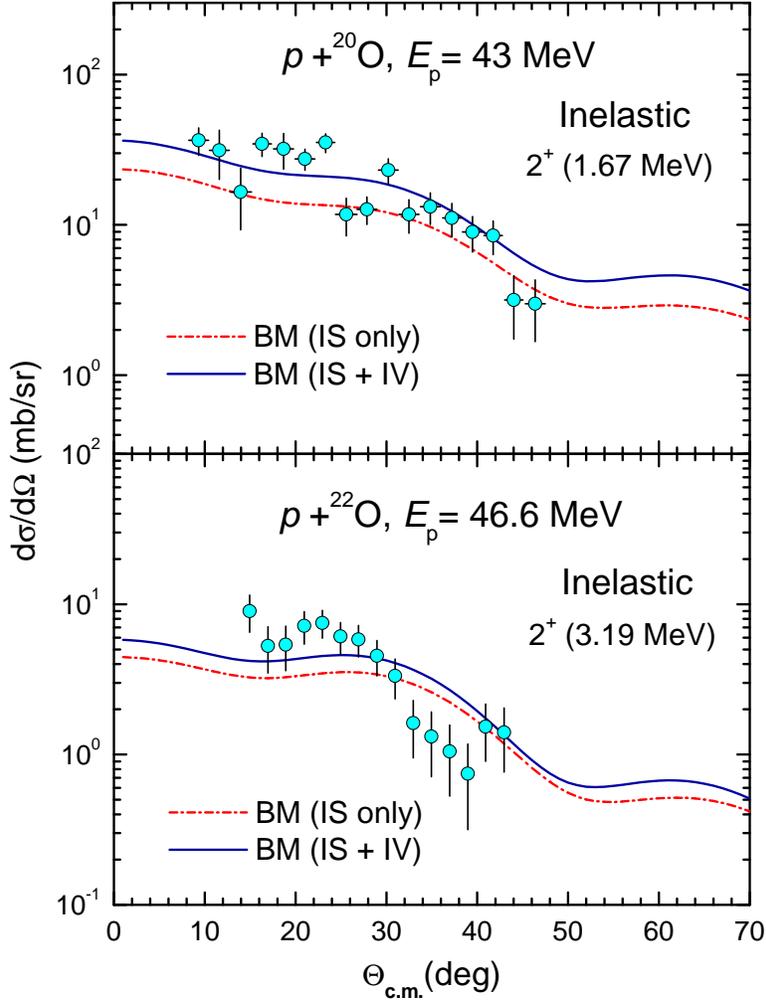,height=15cm}} \vspace*{-0.5cm}
 \caption{(Color online) Inelastic $p+^{20,22}$O scattering data versus the CC
results given by inelastic folded FF with the isospin dependence of the CDM3Y6
interaction included (IS+IV) or neglected (IS only).} \label{f8}
\end{figure}
Given a smaller IV deformation resulted from a weaker core polarization by the
valence neutrons in 2$^+_1$ excitation of $^{22}$O compared to the $^{20}$O
case, it is quite illustrative to show the explicit IV contribution in the
calculated inelastic cross section. We have plotted in Fig.~\ref{f8} the
calculated inelastic $p+^{20,22}$O cross sections given by inelastic folded FF
with the contribution from the isovector part (\ref{p5}) included or neglected.
One can see that the IV contribution is very strong and amounts up 40\% of the
total inelastic cross section in the $^{20}$O case, and it becomes weaker
(around 22\%) in the $^{22}$O case. As can be seen from Eq.~(\ref{p5}), the IV
part of the folded FF is entirely determined by the difference between the
neutron and proton transition densities and its strength is, therefore, directly
proportional to the contribution by the valence neutrons. Consequently, it is
vital to treat the IV dependence of the (effective) $NN$ interaction properly in
a folding model analysis of proton scattering on a neutron-rich target before
the neutron transition strength can be accurately deduced. Although one could
still obtain a good description of the $(p,p')$ data with the IS form factor
only by scaling up the strength of transition densities, like in the folding
model analysis \cite{Gu06} of these same data using the same continuum QRPA
transition densities and \emph{isospin independent} DDM3Y interaction, it is
uncertain to compare the best-fit $M_n/M_p$ ratios obtained in such an analysis
with those deduced by our consistent folding method. We note further that the
more advanced g-folding model \cite{Am01} was also used recently to study the
same inelastic proton scattering data on Oxygens \cite{Kar07}. While the
inelastic cross sections given by this g-folding model agree fairly with data in
the angular shape, the authors need to scale the calculated $(p,p')$ cross
sections by a factor of 2, 5 and 1.6 to fit the data for 2$^+_1$ excitation in
$^{18}$O, $^{20}$O and $^{22}$O, respectively, and this is likely due to a
\emph{truncated} single particle basis based on $0\hbar\omega=0$ shell model
wave functions only. The fact that the largest scaling was needed for the
$^{20}$O case in the g-folding model study also indicates that the higher order
configuration mixing caused by the valence neutrons is strongest in $^{20}$O and
this result is in a sound agreement with our finding. Finally, it is interesting
to note that the contribution by the IV form factor to the $(p,p')$ cross
section for 2$^+_1$ excitation of $^{20}$O is very close to the cross section
shift presumably caused the `isospin impurity' in the $2^+_1$ excitation of the
$A=20,\ T=2$ isobars (compare Fig.~\ref{f6} and upper panel of Fig.~\ref{f8}).

\begin{figure}[htb]
\mbox{\epsfig{file=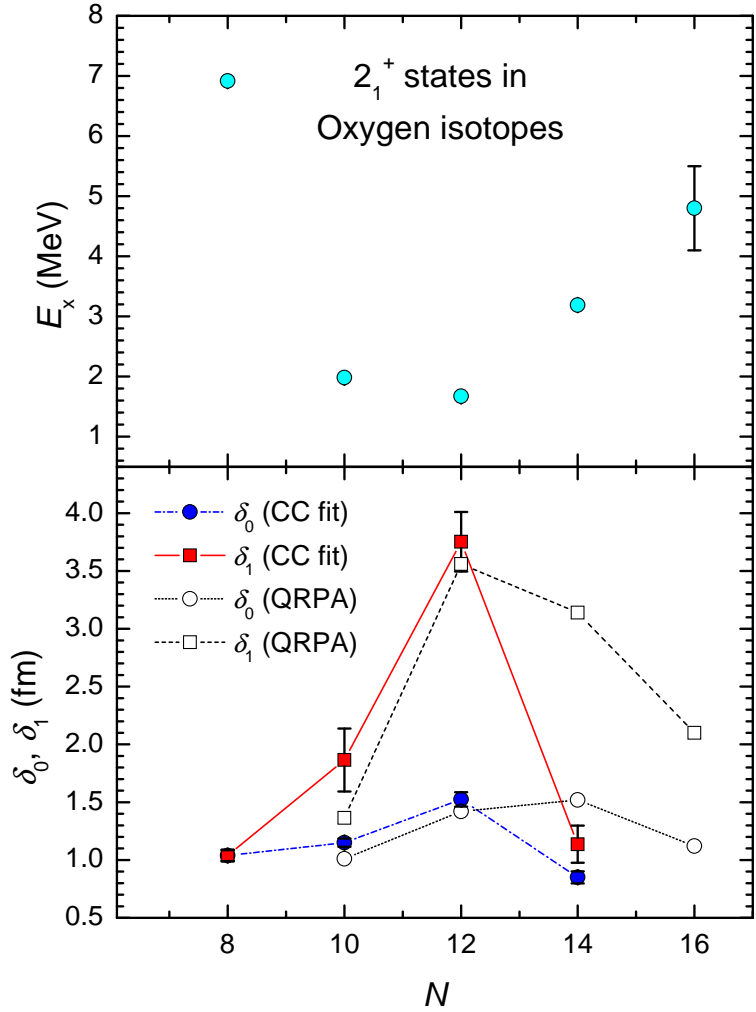,height=15cm}} \vspace*{-0.5cm}
 \caption{(Color online) Observed excitation energies (upper panel), the
isoscalar ($\delta_0$) and isovector ($\delta_1$) deformation lengths (lower
panel) of 2$^+_1$ states in Oxygen isotopes deduced from the folding model
analyses of this work and Ref.~\cite{Kh05a} (for double-closed shell $^{16}$O)
and from the continuum QRPA results \cite{El02}.} \label{f9}
\end{figure}

With a direct connection between the IV deformation and dynamic contribution by
the valence neutrons to the nuclear excitation, it is natural to link the IV
deformation with possible changes of the neutron shell structure. The best-fit
IS and IV deformation lengths of 2$^+_1$ states in Oxygen isotopes and those
derived from the results of continuum QRPA calculation \cite{El02}, using
Eqs.~(\ref{e7}) and (\ref{e7a}), are plotted versus the neutron number $N$ in
Fig.~\ref{f9}. For the double-closed shell $^{16}$O nucleus, we have adopted the
IS limit with $\delta_{0}=\delta_{1}=1.038\pm 0.048$ fm as deduced from the
measured $B(E2)$ value and used in a recent folding model study of inelastic
$^{16}$O+$^{16}$O scattering \cite{Kh05a}.
An enhanced IV deformation (with $\delta_1>\delta_0$) resulted from the core
polarization by the valence neutrons can be seen for the open-shell $^{18,20}$O
nuclei, with maximum of $\delta_1$ observed for 2$^+_1$ state in $^{20}$O or at
$N=12$. Such a maximum of the IV deformation also corresponds to the largest
$M_n/M_p$ ratio found for 2$^+_1$ state in $^{20}$O. With $N$ approaching 14,
the extracted $\delta_1$ value is drastically reduced and becomes rather close
to $\delta_0$ which indicates a much weaker IV mixing in 2$^+_1$ excitation of
$^{22}$O. A similar trend has also been predicted by the continuum QRPA
calculation \cite{El02}, although the predicted difference between $\delta_1$
and $\delta_0$ still remains significant at $N=14$. This difference was
predicted to be substantially smaller at $N=16$ (see open squares and circles in
Fig.~\ref{f9}), and it is natural to suggest from the $N$-dependence shown in
Fig.~\ref{f9} that $\delta_1$ is reaching its second minimum at $N=16$. Based on
a similar $N$ dependence obtained for the IV deformation lengths of 2$^+_1$
states in Sulfur isotopes \cite{Kh07a} where a clear minimum of $\delta_1$ was
found at the neutron magic number $N=20$, the deduced $N$ dependence of
$\delta_1$ for $2^+_1$ states in Oxygen isotopes seems to suggest that the
neutron shell closure occurs again at $N=16$ in Oxygen isotopes. Such a shell
closure scenario is well illustrated by the $N$ dependence of the excitation
energies of 2$^+_1$ states in Oxygen isotopes (upper panel of Fig.~\ref{f9}).
One can see that the energy of 2$^+_1$ state goes through its minimum at $N=12$,
where the core polarizing contribution by the valence neutrons is strongest. As
$N$ moves to $N=16$ this contribution becomes much weaker and the excitation
energy of 2$^+_1$ state becomes larger because of the \emph{enhanced} energy gap
between the 1$s$ and 0$d$ neutron subshells. Although the excitation energy of
2$^+_1$ state in $^{24}$O was predicted by different structure calculations
\cite{El02,Ob05} to be around 4 MeV, the experimental observation has been quite
difficult due to the weak excitation of this state. In particular, no $(p,p')$
data could be measured so far for 2$^+_1$ state in $^{24}$O. An important
evidence has been found recently in the experiment on neutron decay of unstable
Oxygen isotopes by the Michigan State University Group \cite{Fr08,Ho09}, where
2$^+_1$ state of $^{24}$O was identified as a very weak resonance at the
excitation energy of about 4.7 MeV which undergoes direct neutron decay to the
ground state of $^{23}$O. Assuming the weak $E2$ transition strength predicted
by the continuum QRPA \cite{El02} for 2$^+_1$ state in $^{24}$O, our folding
model approach predicts that the $(p,p')$ cross section for this state is at
least factor of 2 smaller than that measured for 2$^+_1$ state in $^{22}$O
\cite{Be06}. The (unbound) excited 5/2$^+$ state of $^{23}$O has also been
observed in the same neutron decay measurement \cite{Fr08} at an excitation
energy of around 2.8 MeV which fits well into the gap of about 4 MeV between the
1$s_{1/2}$ and 0$d_{3/2}$ subshells predicted, e.g., by the
Hartree-Fock-Bogoljubov calculation \cite{Ob05}. Based on this discussion as
well as the systematics on the $\beta$-decay $Q$ values and single neutron
separation energies made by Kanungo {\it et al.} \cite{Ka02} for a wide range of
neutron rich even-even isotopes, we can draw a definitive conclusion on the
neutron shell closure at $N=16$ in unstable Oxygen isotopes. To this end, more
experiments for $^{24}$O, especially, the $(p,p')$ measurement in the inverse
kinematics would be of further interest.

\section{Summary}
A coupled channel analysis of the $^{18,20,22}$O$(p,p')$ scattering data has
been performed, using the OP and inelastic FF calculated microscopically in a
compact folding model approach, to extract the neutron transition matrix
elements $M_n$ as well as the isoscalar ($\delta_0$) and isovector ($\delta_1$)
deformation lengths of 2$^+_1$ states in the Oxygen isotopes, with the proton
transition matrix elements $M_p$ fixed by the measured electric transition rates
$B(E2)$.

The newly determined ratios $M_n/M_p$ for 2$^+_1$ states in $^{18,20}$O have
been compared to those deduced from the isospin symmetry, using the experimental
$B(E2)$ transition rates of 2$^+_1$ states in the proton rich $^{18}$Ne and
$^{20}$Mg isotopes. Given the experimental $B(E2)$ values available for $2^+_1$
states in these four nuclei, a future high-precision $(p,p')$ measurement to
accurately determine the neutron transition strengths of $2^+_1$ states in the
mirror pairs $^{18}$O,$^{18}$Ne and $^{20}$O,$^{20}$Mg (in the inverse
kinematics), should provide vital information on the isospin impurity in the
$2^+_1$ excitation of $A=18,T=1$ and $A=20,T=2$ isobaric multiplets,
respectively.

The enhancement of the IV deformation has been confirmed again for the
open-shell $^{18,20}$O nuclei which show a strong core polarization by the
valence neutrons. Along the isotope chain, the behavior of the dynamic IV
deformation of $2^+_1$ state is closely correlated with the evolution of the
valence neutron shell, and $\delta_1$ has been found to reach its maximum at
$N=12$ which corresponds to the largest $M_n/M_p$ ratio found for 2$^+_1$ state
in $^{20}$O. A fast decrease of the IV deformation towards $N=16$ should be
connected with the neutron shell closure occurring at this new magic number of
neutrons.

\section*{Acknowledgement}
The authors thank Elias Khan for helpful communications on the
$^{18,20,22}$O$(p,p')$ data and QRPA results for the nuclear transition
densities. N.D.C. is grateful to H.S. Than for his assistance in the numerical
calculation. The research was supported, in part, by Natural Science Council of
Vietnam and Vietnam Atomic Energy Commission (VAEC).

\end{document}